\documentclass[aps,prl,twocolumn,superscriptaddress,amsmath,amssymb,showpacs]{revtex4-1}
\usepackage{amsmath}
\usepackage{amssymb}
\usepackage{graphicx}
\usepackage{float}
\usepackage{subfigure}
\usepackage{color}
\usepackage{hyperref}
\usepackage{comment}
\usepackage{makecell}

\begin{document}
\title{Dimensional crossover in Kardar-Parisi-Zhang growth}
\author{Ismael S. S. Carrasco}
\email{ismael.carrasco@unb.br}
\affiliation{International Center of Physics, Institute of Physics, University of Brasilia, 70910-900, Brasilia, Federal District, Brazil}
\author{Tiago J. Oliveira}
\email{tiago@ufv.br}
\affiliation{Departamento de F\'isica, Universidade Federal de Vi\c cosa, 36570-900, Vi\c cosa, MG, Brazil}
\date{\today}

\begin{abstract}
Two-dimensional (2D) KPZ growth is usually investigated on substrates of lateral sizes $L_x=L_y$, so that $L_x$ and the correlation length ($\xi$) are the only relevant lengths determining the scaling behavior. However, in cylindrical geometry, as well as in flat rectangular substrates $L_x \neq L_y$ and, thus, the surfaces can become correlated in a single direction, when $\xi \sim L_x \ll L_y$. From extensive simulations of several KPZ models, we demonstrate that this yields a dimensional crossover in their dynamics, with the roughness scaling as $W \sim t^{\beta_{\text{2D}}}$ for $t \ll t_c$ and $W \sim t^{\beta_{\text{1D}}}$ for $t \gg t_c$, where $t_c \sim L_x^{1/z_{2\text{D}}}$. The height distributions (HDs) also cross over from the 2D flat [cylindrical] HD to the asymptotic Tracy-Widom GOE [GUE] distribution. Moreover, 2D-to-1D crossovers are found also in the asymptotic growth velocity and in the steady state regime of flat systems, where a family of universal HDs exists, interpolating between the 2D and 1D ones as $L_y/L_x$ increases. Importantly, the crossover scalings are fully determined and indicate a possible way to solve 2D KPZ models.
\end{abstract}

\maketitle


Besides a large number of model systems \cite{barabasi,KrugAdv,healy95,Kriecherbauer2010}, one-dimensional (1D) interfaces observed in paper burning fronts \cite{Maunuksela1997,*Miettinen2005}, growth of cell colonies \cite{Huergo10,*Huergo11,*Huergo12}, turbulent phases in liquid crystals \cite{Takeuchi2010,Takeuchi2011} and colloidal deposition at the edges of evaporating drops \cite{Yunker2013a}, as well as 2D surfaces of CdTe \cite{Almeida14,Almeida15,Almeida17}, oligomer \cite{healy14exp} and NiW \cite{Rodolfo17} thin films are all examples of systems whose height field $h(\vec{x},t)$ evolves according to the Kardar-Parisi-Zhang (KPZ) equation \cite{KPZ}
\begin{equation}
 \frac{\partial h(\vec{x},t)}{\partial t} = \nu \nabla^2 h + \frac{\lambda}{2} (\nabla h)^2 + \sqrt{D}\eta(\vec{x},t),
\label{eqKPZ}
\end{equation}
where $\eta$ is a Gaussian white noise, while $\nu$, $\lambda$ and $D$ are phenomenological parameters. A key feature of these systems is the scaling of fluctuations in $h(\vec{x},t)$, usually quantified through the global roughness $W^2=\left< \overline{(h - \bar{h})^2} \right>$, where $\bar{\cdot}$ and $\langle \cdot \rangle$ denote spatial and configurational averages, respectively, at a given time $t$. Indeed, by performing the growth on an initially flat substrate of lateral size $L$, a growth regime (GR) is observed [while the correlation length $\xi \simeq (|\lambda|A^{\frac{1}{2}} t)^{1/z}$ is much smaller than $L$], where $W^2 \simeq \langle\chi^2\rangle_c (\Gamma t)^{2\beta}$. Whereas $\lambda$, $A$ and then $\Gamma \propto |\lambda|A^{\frac{1}{\alpha}}$\footnote{One usually defines $\Gamma = |\lambda|A^2/2$ for 1D KPZ systems \cite{Prahofer2000} and $\Gamma = |\lambda|A^{\frac{1}{\alpha}}$ in higher dimensions \cite{Krug1992,healy12}.} are system-dependent parameters \cite{Krug1992}, the exponents $\alpha$, $\beta$ and $z= \alpha/\beta$, and $\langle\chi^2\rangle_c$ [the second cumulant of the underlying height distribution (HD) $P_{gr}(\chi)$] are universal. The 1D KPZ class is characterized by the exponents $\alpha_{1\text{D}} = 1/2$ and $\beta_{1\text{D}}=1/3$ \cite{KPZ}, and HDs given by Tracy-Widom \cite{TW} distributions from Gaussian orthogonal or unitary ensembles (GOE or GUE) depending respectively on whether the interface is flat (where $L$ is fixed) or circular (where $L \sim t$) \cite{Johansson,Prahofer2000,Sasamoto2010,Amir,Calabrese2011}. Actually, the GUE HD is found whenever $L(t)$ enlarges faster than $\xi(t)$ \cite{Ismael19}. Although the high-dimensional KPZ scenario is far less analytically understood, the exponents $\beta_{d\text{D}} = 7/(8d+13)$ recently found in \cite{tiago22} are very strong candidates to be the exact ones for $d\geq 1$, once they agree strikingly well with numerical estimates up to very large $d$'s \cite{tiago22}. Extensive numerical works have also characterized the properties of the asymptotic GR HDs for flat geometry up to $d=6$ \cite{healy12,healy13,tiago13,Ismael14,Ismael23,Alves14,HHTake2015,JMKim,Alves16}, with the 2D ones being experimentally observed in thin film deposition of different materials \cite{Almeida14,Almeida15,Almeida17,healy14exp}. Beyond the flat case (corresponding to fixed $L=L_x=L_y$), two other main geometries exist for 2D KPZ systems: cylindrical (where $L_x$ is fixed, while $L_y \sim t$) and spherical (where $L=L_x=L_y \sim t$), each one having a characteristic limiting GR HD (for $L_x,L_y \rightarrow \infty$) \cite{healy12,healy13,tiago13,Ismael14,Ismael23}.

At long times, \textit{flat} systems become fully correlated, when $\xi \sim L$, and $W$ saturates at a value that increases as $W_{s}^2 \simeq \langle \zeta^2 \rangle_c A L^{2\alpha}$, where $\langle \zeta^2 \rangle_c$ is the variance of the underlying HD $P_{ssr}(\zeta)$, usually measured in relation to $\bar{h}$ \cite{tiago22av}. Indeed, in this steady state regime (SSR), the HD is different from the GR ones, being Gaussian for 1D \cite{barabasi} and skewed for 2D KPZ systems \cite{Reis2004,Marinari,Pagnani,tiago22av}. The SSR is never observed in the 2D spherical and cylindrical geometries (as well as in the 1D circular one), because at least one side of the substrate enlarges faster than $\xi$. Interestingly, however, cylindrical systems with finite $L_x$ can become fully correlated in the $x$ direction, while $\xi$ keeps increasing forever in the $y$ direction. In practice, this can be realized in the radial growth of 3D cylindrical clusters or the deposition of materials inside V-shaped grooves, of finite length $L_x$ \cite{Ismael23}. A similar scenario may happen for deposition on \textit{flat} rectangular substrates with $L_y \gg L_x$, as is the case, e.g., in selective area growth of nanosheets/nanowalls, horizontal nanowires, etc. \cite{Chi,*Schmid,*Murillo,*Winnerl}, which are appealing for a variety of applications \cite{Yuan,*Wang}.

In this Letter, we present a thorough analysis of this finite-size behavior, unveiling very interesting dimensional crossovers in the roughness scaling, HDs, and growth velocity, as a consequence of the ``single-side-saturation'' of fluctuations. In order to do this, we performed extensive Monte Carlo simulations of three discrete growth models well-known to belong to the KPZ class: the restricted solid-on-solid (RSOS) \cite{KK}, the single step (SS) \cite{ss1} and the Etching model \cite{Mello01}. In all cases, particles are sequentially deposited at random positions of rectangular square lattice substrates, with $L_x\times L_y$ sites and periodic boundary conditions in both directions. Once a site $(i,j)$, with first neighbors $\partial_{ij}$, is sorted, the aggregation rules there are as follows: \textit{RSOS}: $h_{ij} \rightarrow h_{ij} +1$ if $|h_{ij} - h_{k}| \leq 1$ after deposition $\forall$ $k \in \partial_{ij}$; \textit{SS}: $h_{ij} \rightarrow h_{ij} + 2$ if $|h_{ij} - h_{k}| = 1$ after deposition $\forall$ $k \in \partial_{ij}$; \textit{Etching}: $h_{k} = \max[h_{ij}, h_{k}]$  $\forall$ $k \in \partial_{ij}$ and then $h_{ij} \rightarrow h_{ij} +1$. In the RSOS and SS models, the particle is rejected whenever the restrictions above are not satisfied. A checkerboard initial condition, $h_{ij}(t=0)=[1+(-1)^{i+j}]/2$, is used in the SS model, while $h_{ij}(t=0)=0$ in the other cases. In flat geometry, $L_x$ and $L_y$ stay fixed during the growth. To investigate the cylindrical geometry, $L_x$ is kept fixed while $L_y$ enlarges as $\langle L_y \rangle = L_0 + \omega t$ by random duplications of columns in the $y$ direction \cite{Ismael23}. Such duplications occur with probability $p_e = \omega/(L_x L_y + \omega)$ and are stochastically mixed with particle deposition (occurring with probability $1-p_e$). After each of these events, the time is increased by $\Delta t = 1/(L_x L_y + \omega)$, with $\omega=0$ in the flat case \footnote{For more details on this method, see, e.g., Refs. \cite{Ismael14,Ismael23,Ismael22}.}. Results for several $(L_x,L_y)$ will be presented below, with $L_0 =\omega=2$ in cylindrical geometry.

\begin{figure}[t]
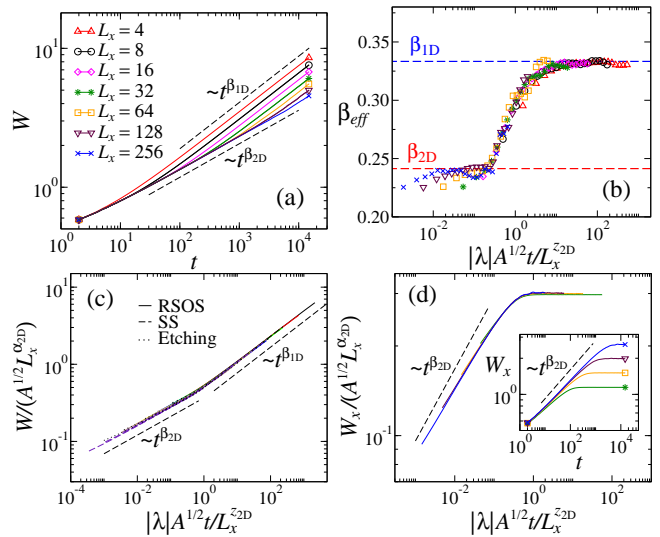

 \includegraphics[width=4.2cm]{Fig1a.eps}
 \includegraphics[width=4.2cm]{Fig1b.eps}
 \includegraphics[width=4.2cm]{Fig1c.eps}
 \includegraphics[width=4.2cm]{Fig1d.eps}
 \caption{(a) Roughness $W$ versus time $t$. (b) Effective growth exponents $\beta_{eff}$, (c) rescaled roughness $W/(A^{\frac{1}{2}}L_x^{\alpha_{2\text{D}}})$ and (d) rescaled ``line roughness'' $W_x/(A^{\frac{1}{2}}L_x^{\alpha_{2\text{D}}})$ against $t/t_c$. The insertion in (d) shows the non-rescaled data for $W_x$. The dashed lines have the indicated slopes. Results for the RSOS model on flat substrates and several $L_x$ are shown in each panel. Data for the other models are also depicted in (c). Panel (a) shows curves for both $L_y=32768$ and 65536, which have a negligible difference for a given $L_x$, demonstrating that $L_y$ is irrelevant in this crossover (provided that $L_y \gg L_x$).}
\label{fig1}
\end{figure}

{\bf \textit{Growth regime}.} We start our study focusing on the GR, which is the single regime observed in cylindrical geometry. To keep $\xi \ll L_y$ in the flat case, $L_y = 32768$ was used in most of this analysis, but some results for $L_y = 65536$ will be also shown in some cases, when indicated. Figure \ref{fig1}(a) presents the temporal variation of the roughness $W$ for the RSOS model in flat geometry. Interestingly, it displays a crossover from an initial 2D regime, where $W \sim t^{\beta_{2\text{D}}}$, to an asymptotic 1D scaling, with $W \sim t^{\beta_{1\text{D}}}$. The very same behavior is found for all models, in both geometries, as demonstrated in the Suppl. Mat. (SM) \cite{SM}. This crossover is better appreciated in the effective growth exponents $\beta_{eff}(t)$ --- the successive slopes in curves of $\ln W \times \ln t$ ---, which clearly change from $\beta_{2\text{D}} \approx 0.2414$ \cite{Kelling2018,tiago22} to $\beta_{1\text{D}}=1/3$ as time evolves [see Fig. \ref{fig1}(b)]. The crossover time, $t_c$, increases with $L_x$, as observed in Fig. \ref{fig1}(a), and the same obviously happens in the curves of $\beta_{eff} \times t$. Notwithstanding, when the time is rescaled by $t_c = L_x^{z_{2\text{D}}}/(|\lambda|A^{\frac{1}{2}})$ \footnote{Since $z = \alpha/\beta$ and $\alpha + z = 2$ (for KPZ systems) \cite{barabasi}, $\beta_{2\text{D}} \approx 0.2414$ \cite{Kelling2018,tiago22} means that $\alpha_{2\text{D}} \approx 0.3889$ and $z_{2\text{D}} \approx 1.6111$.}, with $\lambda$ and $A$ assuming their values for the 2D models (given in Tab. 1 of Ref. \cite{Ismael14}), data for different $L_x$ collapse onto a single curve, as observed in Fig. \ref{fig1}(b). An almost perfect collapse is found also in the roughness of all models versus $t/t_c$, when $W$ is rescaled by $A^{\frac{1}{2}}L_x^{\alpha_{2\text{D}}}$ [see Fig. \ref{fig1}(c)]. Hence, the dimensional crossover scaling is given by
\begin{equation}
 W(L_x,t) \simeq A^{\frac{1}{2}} L_x^{\alpha_{2\text{D}}} \mathcal{F}_c\left(\frac{|\lambda|A^{\frac{1}{2}}t}{L_x^{z_{2\text{D}}}}\right),
 \label{CrossScaling}
\end{equation}
for $L_y \gg L_x$, with an universal scaling function
\[
\mathcal{F}_c(x) \sim
\begin{cases}
  x^{\beta_{2\text{D}}}, & \text{for} \quad x \ll 1; \\
  x^{\beta_{1\text{D}}}, & \text{for} \quad x \gg 1.
\end{cases} 
\]
Note that this is analogous to the Family-Vicsek (FV) scaling \cite{FV} (for flat 2D KPZ systems with $L_x=L_y$), with the exception that $\mathcal{F}_{FV}(x) \sim const.$ for $x \gg 1$. This saturation regime will be discussed later on here.

\begin{figure}[t]
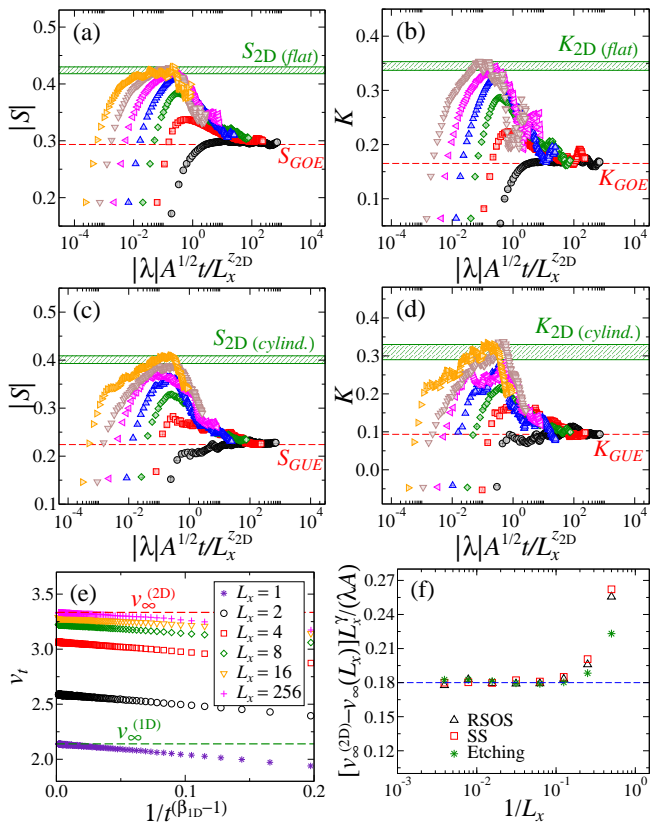

 \includegraphics[width=4.2cm]{Fig2a.eps}
 \includegraphics[width=4.2cm]{Fig2b.eps}
 \includegraphics[width=4.2cm]{Fig2c.eps}
 \includegraphics[width=4.2cm]{Fig2d.eps}
 \includegraphics[width=4.2cm]{Fig2e.eps}
 \includegraphics[width=4.2cm]{Fig2f.eps}
 \caption{GR HDs' skewness $|S|$ and kurtosis $K$ against $t/t_c$, for the RSOS model in flat [(a) and (b)] and cylindrical [(c) and (d)] geometries, with $L_x =4,8,\ldots,256$, increasing from right-to-left. The values of $S_{GOE/GUE}$ and $K_{GOE/GUE}$ are given, e.g., in Ref. \cite{Prahofer2000}, while the estimates for the 2D HDs were extracted from Ref. \cite{Ismael23}. (e) Extrapolations of the growth velocities $v_t$ to $t\rightarrow \infty$, for the Etching model; the indicated values of $v_{\infty}$ for 1D and \textit{square} 2D substrates are those reported in Ref. \cite{Ismael14}. (f) Rescaled asymptotic velocities $[v_{\infty}^{(2\text{D})}-v_{\infty}(L_x)]L_x^{\gamma}/(\lambda A)$ versus $1/L_x$, with $\gamma=(2-2\alpha_{2\text{D}})$.}
\label{fig2}
\end{figure}

The $t_c$ found above demonstrates that the crossover happens when $\xi \approx L_x$, so that $(|\lambda|A^{\frac{1}{2}} t_c)^{1/z_{2\text{D}}} \approx L_x$. To deeply understand how this single-side correlation affects the KPZ growth, we investigate also the ``line roughness'' $W_{\ell}^2 = \left< \overline{(h - \bar{h})^2} \right>_{\ell}$, where $\bar{\cdot}$ is calculated along lines in the $\ell(=x,y)$ direction. When $L_x=L_y$, both directions are equivalent, leading to $W_x = W_y\approx W$. For the systems analyzed here, one also finds $W_x \approx W_y\approx W \sim t^{\beta_{2\text{D}}}$ at short times. However, for $t \gg t_c$, $W_x$ saturates [following the FV scaling, as seen in Fig. \ref{fig1}(d)], while $W_y$ keeps augmenting as $W_y \approx W \sim t^{\beta_{1\text{D}}}$. Thereby, for $t \gtrsim t_c$ the height fluctuations (and $\xi$) only increase in the $y$ direction, leading the system to scale as if it was 1D.

Figures \ref{fig2}(a)-(d) show the temporal evolution of the skewness $S=\langle h^{3}\rangle_c/\langle h^{2}\rangle_c^{3/2}$ and (excess) kurtosis $K=\langle h^{4}\rangle_c/\langle h^{2}\rangle_c^{2}$ of the GR HDs for the RSOS model in flat and cylindrical geometries. (Here, $\langle h^{n}\rangle_c$ is the $n$th cumulant of the one-point HD.) In both cases, as $L_x$ increases one sees these ratios converging to the values for the respective 2D KPZ HDs at short times. At $t \gtrsim t_c$, however, they start decreasing towards the ratios for the 1D KPZ HDs, with the 2D flat (cylindrical) HD giving place to an asymptotic TW-GOE (TW-GUE) distribution. Hence, not only the roughness scaling but also the underlying HDs undergo a 2D-to-1D crossover.

Similarly to the ``line roughness'', we can also define the ``line HDs'' $P_x(h)$ and $P_y(h)$, measured along lines in the $x$- and $y$-directions, respectively. The $n$th moment of $P_{\ell}(h)$ is given by $M_n^{(\ell)} = \left< \overline{(h - \bar{h})^n} \right>_{\ell}$, from which the skewness $S_{\ell}=M_3^{(\ell)}/W_{\ell}^3$ and (excess) kurtosis $K_{\ell}=M_4^{(\ell)}/W_{\ell}^4-3$ are calculated. As demonstrated in the SM \cite{SM}, $|S_y|$ and $K_y$ behave analogously to $|S|$ and $K$ in Figs. \ref{fig2}(a)-(d). At short times, curves of $|S_x|$ and $K_x$ versus $t$ also display maxima that converge to the ratios of the 2D KPZ GR HDs, but present stronger finite-size corrections than $|S_y|$ and $K_y$, as expected. Remarkably, $S_x$ and $K_x$ saturate at long times, instead of displaying slow crossovers toward the 1D HDs. The saturation values $|S_{x,s}| \approx K_{x,s} \approx 0.11$ are found for all models in both geometries (see \cite{SM}). Since they do not fit in any known distribution for 1D or 2D KPZ systems, a new universal KPZ HD exists at the single-side-saturation regime.

According to the ``KPZ ansatz'' for the one-point height [$h \simeq v_{\infty} t + (\Gamma t)^{\beta}\chi$], the asymptotic growth velocity $v_{\infty}$ can be obtained by extrapolating $v_t = \partial_t \left< \bar{h} \right>$ versus $t^{\beta_{1\text{D}}-1}$ to $t \rightarrow \infty$. Examples of such extrapolations (for the Etching model) are shown in Fig. \ref{fig2}(e), demonstrating that $v_{\infty}$ is a function of $L_x$, which smoothly changes from $v_{\infty}^{(1\text{D})}$ to $v_{\infty}^{(2\text{D})}$ as $L_x$ increases. A similar behavior is found for the RSOS and SS models, but with $v_{\infty}$ decreasing with $L_x$, because $v_{\infty}^{(1\text{D})} > v_{\infty}^{(2\text{D})}$ for these systems (see the SM \cite{SM}). Substantially, while the asymptotic growth exponent and HDs are those for the 1D KPZ class, the growth velocity can assume any value between $v_{\infty}^{(1\text{D})}$ and $v_{\infty}^{(2\text{D})}$. To explain this behavior, we recall that on a flat 2D substrate of lateral size $L$, $v_t$ saturates (in the SSR) at a value $v(L) \simeq v_{\infty}^{(2\text{D})} - (\lambda A/2) L^{2\alpha_{2\text{D}}-2}$ \cite{Krug90,Krug1992}. Hence, for the single-side-saturation observed here, it is reasonable to expect that
\begin{equation}
v_{\infty}(L_x) \simeq v_{\infty}^{(2\text{D})} - c \lambda A L_x^{-(2-2\alpha_{2\text{D}})}.
\end{equation}
This ansatz is confirmed in Fig. \ref{fig2}(f), which shows that $[v_{\infty}^{(2\text{D})}-v_{\infty}(L_x)]L_x^{2-2\alpha_{2\text{D}}}/(\lambda A)$ does indeed converge to a universal constant value $c \approx 0.18$, for large $L_x$, when $\lambda$ and $A$ assume their values for the 2D models \cite{Ismael14}.

\begin{figure}[t]
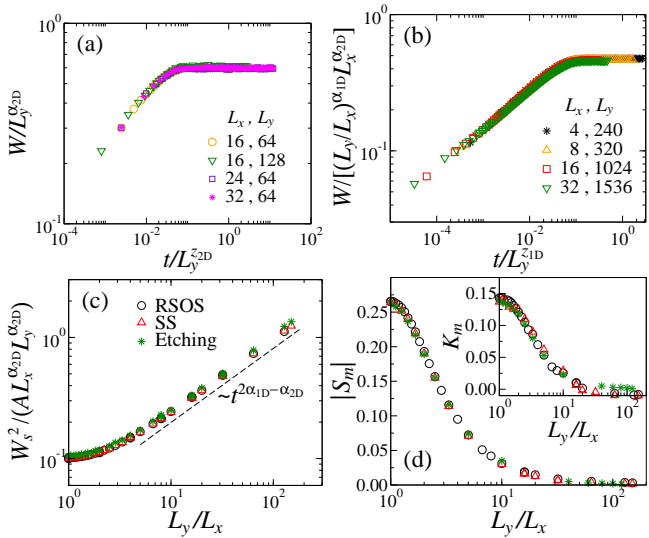

 \includegraphics[width=4.2cm]{Fig3a.eps}
 \includegraphics[width=4.2cm]{Fig3b.eps}
 \includegraphics[width=4.2cm]{Fig3c.eps}
 \includegraphics[width=4.2cm]{Fig3d.eps}
 \caption{Rescaled curves of roughness $W$ versus time $t$ for the Etching model on flat substrates with small (a) and large (b) aspect ratios $L_y/L_x$. (c) Rescaled squared saturation roughness $W_s^2/(A L_x^{\alpha_{2\text{D}}} L_y^{\alpha_{2\text{D}}})$ against $L_y/L_x$. (d) Variation of the SSR HDs' skewness $S_m$ (main plot) and kurtosis $K_m$ (insertion) with $L_y/L_x$. Results for the three investigated models are shown in (c) and (d), as indicated by the legend in (c).}
\label{fig3}
\end{figure}

{\bf \textit{Steady state regime}.} At long times, the \textit{flat} surfaces become correlated also in the $y$ direction, when $\xi \sim L_y > L_x$, and the SSR is attained, as confirmed in Figs. \ref{fig3}(a) and \ref{fig3}(b). As expected, the saturation time $t_s$ (where the crossover to the SSR takes place) is an increasing function of $L_y$, but it behaves differently for small and large aspect ratios $L_y/L_x$. Indeed, rescaled curves of $W/L_y^{\alpha_{2\text{D}}}$ versus $t/L_y^{z_{2\text{D}}}$ collapse reasonably well for $L_y/L_x \lesssim 8$, as shows Fig. \ref{fig3}(a), indicating that $t_s \sim L_y^{z_{2\text{D}}}$ and, thus, the FV scaling holds when $L_y/L_x$ is small. On the other hand, one sees in Fig. \ref{fig3}(b) that the data for large $L_y/L_x$ collapse when $W$ is rescaled by $(L_y/L_x)^{\alpha_{1\text{D}}} L_x^{\alpha_{2\text{D}}}$, with $t$ divided by $L_y^{z_{1\text{D}}}$, so that now $t_s \sim L_y^{z_{1\text{D}}}$. These behaviors suggest that, for a given $L_y/L_x$, the saturation roughness, $W_s$, scales with the effective linear size $\bar{L} = \sqrt{L_x L_y}$ as $W_s \sim A^{\frac{1}{2}} \bar{L}^{\alpha_{2\text{D}}}$. In fact, $W_s^2/(A \bar{L}^{2\alpha_{2\text{D}}})$ versus $L_y/L_x$,  for different models and sizes, fall into a universal crossover curve [see Fig. \ref{fig3}(c)], indicating that
\begin{equation}
 W_s^2 \simeq A L_x^{\alpha_{2\text{D}}} L_y^{\alpha_{2\text{D}}} \mathcal{G}\left( \frac{L_x}{L_y} \right),
 \label{ScalingSS}
\end{equation}
where, once again, the parameters $A$ for the 2D models have to be used. The scaling function $\mathcal{G}(y)$ behaves as $\mathcal{G}(y) \approx const. \approx 0.10$ for $y \sim 1$. This is indeed expected, since for $y=1$ one recovers the usual 2D scaling: $W_{s}^2 \simeq \langle \zeta^2 \rangle_c A L^{2\alpha_{2\text{D}}}$, with $\langle \zeta^2 \rangle_c \approx 0.103$ \cite{tiago22av}. For $y \gg 1$, we find strong evidence that $\mathcal{G}(y) \sim y^{2\alpha_{1\text{D}}-\alpha_{2\text{D}}}$, meaning that $W_s \simeq B L_y^{\alpha_{1\text{D}}}$ with $B \sim L_x^{-(\alpha_{1\text{D}}-\alpha_{2\text{D}})}$. Namely, when $L_y \gg L_x$, the system displays the 1D scaling in $L_y$, but with a $L_x$-dependent scaling amplitude.

Figure \ref{fig3}(d) shows the variation of the asymptotic values for the skewness $S_m=M_3/W_s^3$ (main plot) and kurtosis $K_m=M_4/W_s^4-3$ (insertion) for the SSR HDs as functions of $L_y/L_x$. (Here, $M_n = \left< \overline{(h - \bar{h})^n} \right>$ is $n$th moment of the SSR HDs.) The data collapse confirms that $L_y/L_x$ is indeed the relevant quantity determining the behavior in the SSR. Moreover, this makes it clear that universal curves of $S(L_y/L_x)$ and $K(L_y/L_x)$ exist, which start at the 2D values for the moment ratios for $L_y/L_x = 1$, decrease as $L_y/L_x$ augments and finally vanish when $L_y/L_x \rightarrow \infty$. Therefore, in this limit, the SSR HD is Gaussian, as expected for 1D KPZ systems.

We conclude recalling that crossovers from an initial Edwards-Wilkinson dynamics (given by Eq. \ref{eqKPZ} with $\lambda=0$ \cite{EW}) to an asymptotic KPZ scaling, as well as from random (Eq. \ref{eqKPZ} with $\nu=\lambda=0$ \cite{barabasi}) to KPZ growth have been widely investigated in past decades \cite{[{See, e.g., }]Daryaei,*Alice,*Hedayatifar,*Juvenil,*tiago13EWKPZ,*[{ and references therein}]}, besides some other forms of temporal crossovers \cite{barabasi,Almeida17,TakeuchiCross}. However, to the best of our knowledge, the dimensional crossover found here has never been observed so far. However, for square substrates, as usually considered in previous works on 2D KPZ growth, $t_s = t_c$ and no dimensional crossover is observed, because the SSR precedes the raising of the 1D scaling. On the other hand, when $L_y \gg L_x$ one has $t_s \gg t_c$ and, then, the 1D regime is observed for $t_c \ll t \ll t_s$. This strongly suggests that the fate of any 2D KPZ cylindrical system with finite $L_x$ (where $L_y$ increases indefinitely, so that $t_s \rightarrow \infty$) is to follow the 1D circular KPZ subclass, with $W \sim t^{1/3}$ and GUE HD, at long times. Flat 2D KPZ systems with finite $L_x$ and $L_y \rightarrow \infty$ will fall asymptotically in the 1D flat KPZ subclass, with $W \sim t^{1/3}$ and GOE HD. When $L_y$ $(> L_x)$ is also finite, a SSR is attained at long times, where both the scaling behavior and HDs depend on the aspect ratio $L_y/L_x$. Interestingly, while the height fluctuations select one of the few universal KPZ HDs in the GR, in the SSR we find a continuous family of distributions, demonstrating that the SSR exhibits a less robust behavior than the GR. Beyond representing a substantial generalization of the current knowledge on interface growth, these results indicate that rectangular substrates may be an important platform for theoretical studies of KPZ systems. In fact, since their 1D asymptotic behaviors are analytically known, this may be an interesting starting point to obtain the long-waited exact solutions of the 2D regime. We also notice that in higher dimensions --- let us say, for a $d$-dimensional flat substrate with lateral sizes $1 \ll L_1 \ll L_2 \ll \cdots \ll L_d$ --- a sequence of temporal crossovers are expected in the GR, from an initial $W \sim t^{\beta_{dD}}$ behavior to $W \sim t^{\beta_{(d-1)D}}$ and so on, until the final $W \sim t^{\beta_{1D}}$ scaling. It turns out that, already in the 3D case, this interesting scenario may be difficult to observe in simulations, since it is quite hard to perform the deposition on substrates with $1 \ll L_1 \ll L_2 \ll L_3$ for long growth times. From a more applied perspective, our work provides important insights into the selective growth of nanostructures on stripe patterns, whose width can be as small as $L_x \sim 100 nm$, while $L_y \sim 1 \mu m$ \cite{Yuan}. Even though their growth dynamics may likely differ from the KPZ one, if the deposition time is large enough to yield $\xi \gg L_x$, a single-side-saturation --- with the consequent dimensional crossover, as unveiled here --- can be expected in any type of correlated growth.

\acknowledgments

The authors acknowledge partial financial support from CNPq, FAPEMIG, and FAPDF (grant number 00193-00001817/2023-43) (Brazilian agencies); and thank C. I. L. Ara\'ujo for helpful discussions.

\bibliography{bibExpKPZ2D}

\clearpage

\appendix


\onecolumngrid

\section{Supplemental material}

\subsection{Additional results for the dimensional crossover in the roughness scaling}
\label{secWxt}

\begin{figure*}[h!]
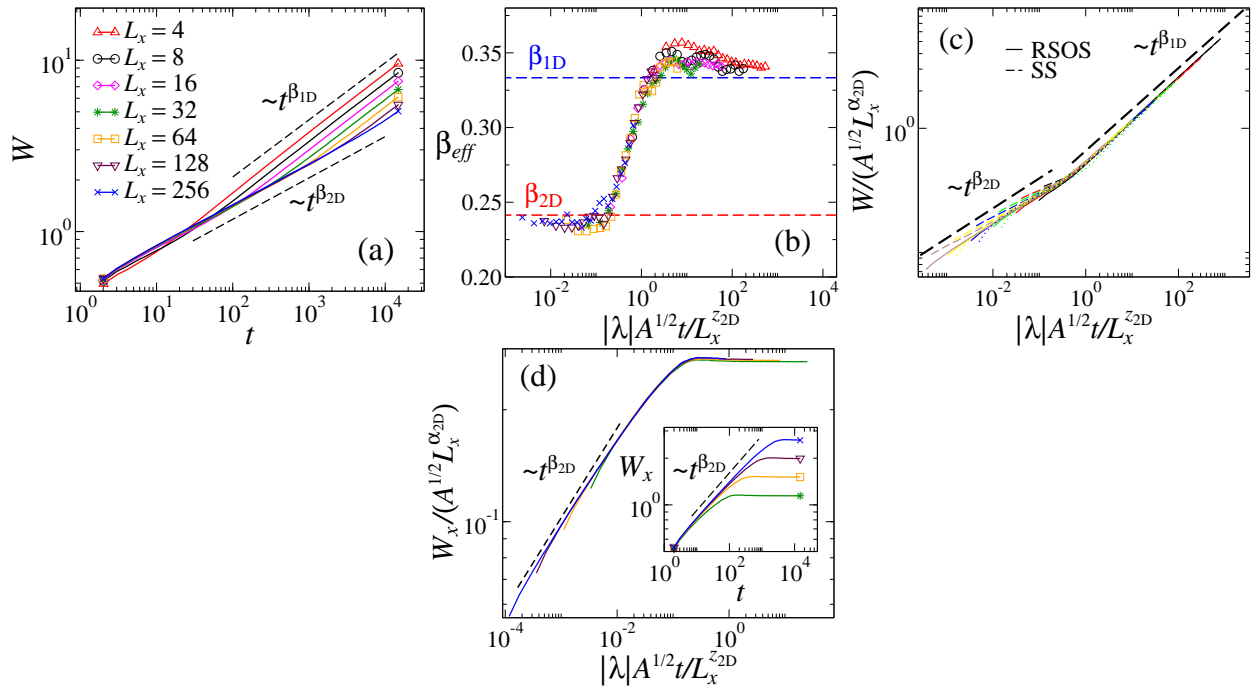

	\includegraphics[height=4.5cm]{FigSM1a.eps}
	\includegraphics[height=4.5cm]{FigSM1b.eps}
	\includegraphics[height=4.5cm]{FigSM1c.eps}
	\includegraphics[height=4.5cm]{FigSM1d.eps}
	\caption{(a) Roughness $W$ versus time $t$. (b) Effective growth exponents $\beta_{eff}$, (c) rescaled roughness $W/(A^{\frac{1}{2}}L_x^{\alpha_{2\text{D}}})$ and (d) rescaled ``line roughness'' $W_x/(A^{\frac{1}{2}}L_x^{\alpha_{2\text{D}}})$ against $t/t_c$, where $t_c = L_x^{z_{2\text{D}}}/(|\lambda|A^{\frac{1}{2}})$. The insertion in (d) shows the non-rescaled data for $W_x$. The dashed lines have the indicated slopes. Results for the RSOS model in cylindrical geometry and several $L_x$ are shown in each panel. Data for the other models are also depicted in (c).}
	\label{fig1SM}
\end{figure*}

The results presented in Fig. 1 of the main text are for the KPZ models in flat geometry. Analogous data for cylindrical geometry are depicted in Fig. \ref{fig1SM} here, where the very same behavior of the flat case is found, but with stronger finite-size and -time corrections. This confirms that the dimensional crossover is a general feature of 2D KPZ systems with $L_x \ll L_y$, where the surfaces can become correlated in a single direction. In other words, in spite of what is happening in the $y$ direction (i.e., of whether it is expanding or not), when $\xi \approx L_x$ the fluctuations saturate in the $x$ direction, yielding the dimensional crossover in the surface roughness, following the scaling behavior given by Eq. 2 of the main text.

\subsection{Directional height fluctuations}
\label{secLineHDs}

\begin{figure*}[h!]
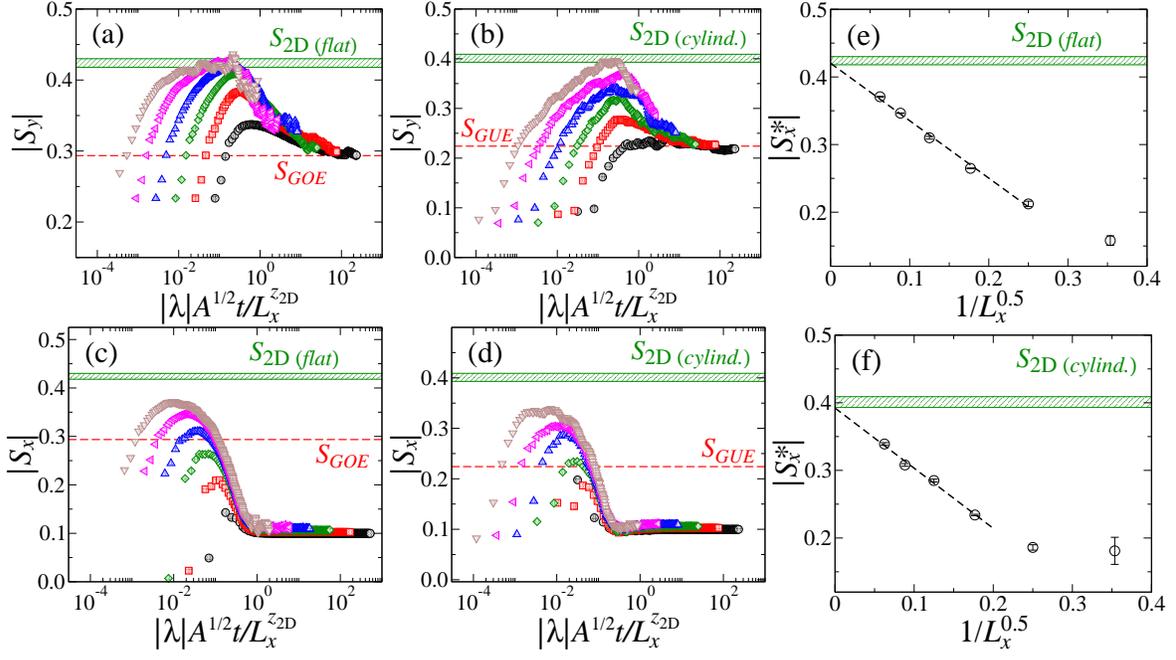

	\includegraphics[height=4.3cm]{FigSM2a.eps}
	\includegraphics[height=4.3cm]{FigSM2b.eps}
	\includegraphics[height=4.3cm]{FigSM2e.eps}
	\includegraphics[height=4.3cm]{FigSM2c.eps}
	\includegraphics[height=4.3cm]{FigSM2d.eps}
	\includegraphics[height=4.3cm]{FigSM2f.eps}
	\caption{Temporal evolution of the skewness of the ``line HDs'' for the $y$- [(a) and (b)] and $x$-direction [(c) and (d)], for the RSOS model in flat [(a) and (c)] and cylindrical [(b) and (d)] geometries. The values of the maxima ($|S_x^*|$) in the curves of $|S_x|$ for flat and cylindrical geometries are extrapolated to the $L_x \rightarrow \infty$ limit in panels (e) and (f), respectively.}
	\label{fig2SM}
\end{figure*}

\begin{figure*}[h!t]
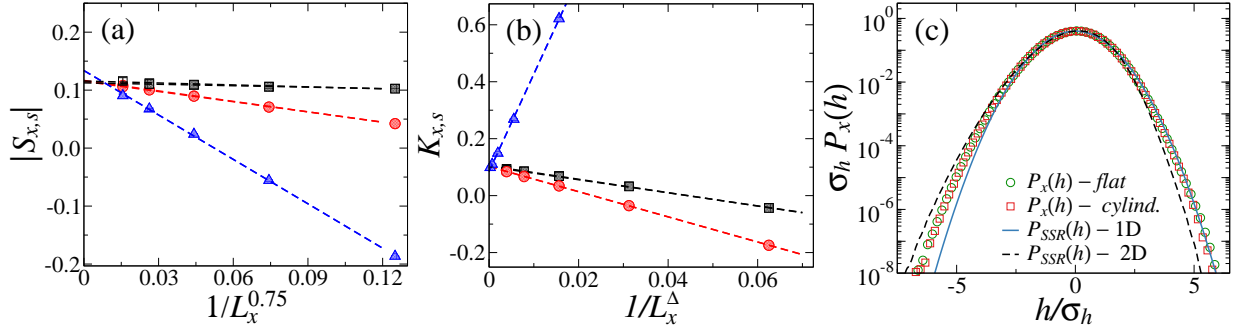

	\includegraphics[height=4.3cm]{FigSM3a.eps}
	\includegraphics[height=4.3cm]{FigSM3b.eps}
	\includegraphics[height=4.3cm]{FigSM3c.eps}
	\caption{Extrapolations to $L_x\rightarrow \infty$ of the saturation values of the (a) skewness $|S_{x,s}|$ and (b) kurtosis $K_{x,s}$ for the RSOS (black squares), SS (red circles), and Etching (blue triangles) in flat (open) and cylindrical (full symbols) geometries. In (b), $\Delta=1$ for RSOS and SS models, while $\Delta=1.5$ in the Etching case. (c) Comparison of the distribution $P_x(h)$ (rescaled by the standard deviation $\sigma_h$) for the single-side-saturation regime (symbols) with those for the steady state regime, $P_{ssr}(h)$, of 1D and 2D KPZ systems.}
	\label{fig3SM}
\end{figure*}

As discussed in the main text, since the ``line roughness'' $W_x$ behaves differently from $W_y$ in the systems we are investigating, it is interesting to analyze also the height distributions (HDs) measured along lines in the $x$ and $y$ directions. Namely, we determine the ``line HDs'' $P_x(h)$ and $P_y(h)$, measured about the mean height of each \textit{line} ($\bar{h}$), such that $P_{\ell}(h)dh$ gives the probability of finding a height (relative to $\bar{h}$) in the interval $[h,h+dh]$, for $\ell =x$ or $y$. 

Fig. \ref{fig2SM}(a) and \ref{fig2SM}(b) display the temporal variation of the skewness $S_y$ (defined in the main text) for the RSOS model in flat and cylindrical geometries, respectively. Notably, its behavior is the same found in Fig. 2 of the main text for the ``global'' skewness; namely, at short times $|S_y|$ converges to the values of the respective 2D KPZ GR HDs and then it displays a slow crossover towards $S_{GOE}$ or $S_{GUE}$ (i.e., the values of the 1D KPZ GR HDs for flat or circular geometries). Analogous results (not shown) are found for the kurtosis $K_y$, which also behaves similarly to $K$ in Fig. 2 of the main text.

The curves of $|S_x|$ versus $t$, for the RSOS model in flat and cylindrical cases, are depicted in Figs. \ref{fig2SM}(c) and \ref{fig2SM}(d), respectively. As we may see, they also have a maximum at short times, but display a slower convergence than that observed in $S_y$, since $L_x \ll L_y$. Thereby, to confirm that such maxima (let us denote them by $S_x^*$) do indeed converge to the expected values for the respective 2D KPZ GR HDs of each geometry, we have extrapolated them to the $L_x \rightarrow \infty$, as shown in Figs. \ref{fig2SM}(e) and \ref{fig2SM}(f). Once again, an analogous behavior (not shown) is found for the kurtosis $K_x$ at short times.

Interestingly, both $S_x$ and $K_x$ saturate when the surfaces become correlated in the $x$-direction, while $S_y$ and $K_y$ present the dimensional crossover. As seen in Figs. \ref{fig2SM}(c) and \ref{fig2SM}(d), the saturation values ($|S_{x,s}|$) for the RSOS model have a mild dependence on $L_x$. For the other models, however, stronger finite-size corrections exist in both $|S_{x,s}|$ and $K_{x,s}$, but, extrapolating these ratios to the $L_x \rightarrow \infty$ limit, we find clear evidence that they converge to the same value $|S_{x,s}| \approx K_{x,s} \approx 0.11$, for both flat and cylindrical geometries [see Figs. \ref{fig3SM}(a) and \ref{fig3SM}(b)]. Therefore, a single HD $P_x(h)$ exists at the single-side-saturation regime, whose shape is shown in Fig. \ref{fig3SM}(c). Interestingly, this HD is different from the one for the SSR of 1D KPZ systems (which is Gaussian, having thus $S=K=0$) and 2D KPZ systems, for which $|S|\approx 0.26$ and $K \approx 0.13$ \cite{tiago22}.

\subsection{Additional results for the asymptotic growth velocity}
\label{secVelocity}

\begin{figure*}[h!]
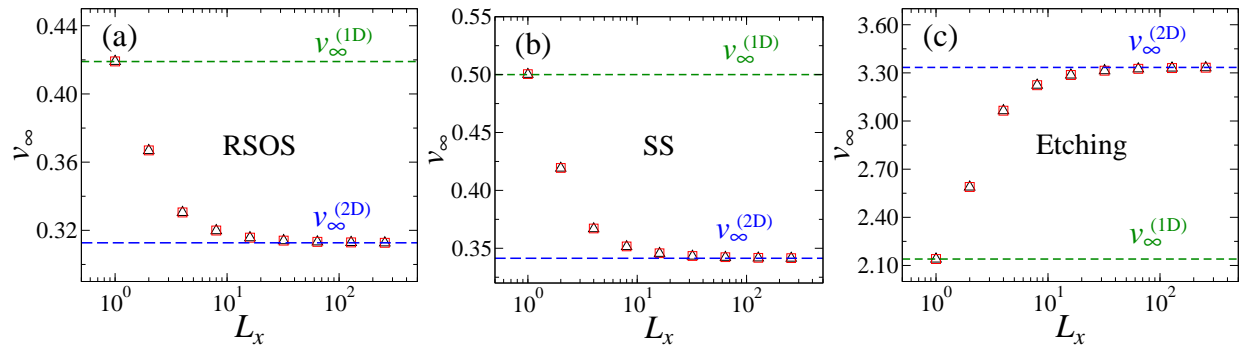

	\includegraphics[height=4.5cm]{FigSM4a.eps}
	\includegraphics[height=4.5cm]{FigSM4b.eps}
	\includegraphics[height=4.5cm]{FigSM4c.eps}
	\caption{Asymptotic growth velocity $v_{\infty}$ versus $L_x$ for the (a) RSOS, (b) SS, and (c) Etching models. Results for flat (green triangles) and cylindrical (red squares) geometries are shown. The values of $v_{\infty}$ for these models on 1D and 2D substrates (with $L_x=L_y$) were extracted from Ref. \cite{Ismael14} and are indicated by the dashed lines in each panel.}
	\label{fig4SM}
\end{figure*}

From extrapolations of $v_t$ versus $t^{\beta_{1\text{D}}-1}$ to $t \rightarrow \infty$ --- as those presented in Fig. 2(e) of the main text for the Etching model and similar ones for the other models, in both geometries ---, we obtain the asymptotic growth velocities $v_{\infty}(L_x)$ for the $L_y \rightarrow \infty$ limit. These asymptotic velocities are shown in Fig. \ref{fig4SM} here, where one may see that, as $L_x$ increases, $v_{\infty}$ presents a smooth variation between the 1D and 2D values found in the literature for these models (given, e.g., in Ref. \cite{Ismael14}). It is also noteworthy that, for a given model, $v_{\infty}(L_x)$ is the same (within the error bars) for flat and cylindrical geometries. This is in full agreement with previous studies demonstrating that the growth velocity is not affected by the substrate expansion in deposition on enlarging substrates \cite{Ismael14,Ismael22,Ismael23}.

\end{document}